\documentstyle[12pt,epsfig]{article}

\def\appendix{{\newpage\section*{Appendix}}\let\appendix\section%
        {\setcounter{section}{0}
        \gdef\thesection{\Alph{section}}}\section}

\newcommand{\non}{\nonumber \\}
\newcommand{\beq}{\begin{equation}}
\newcommand{\eeq}{\end{equation}}
\newcommand{\beqar}{\begin{eqnarray}}
\newcommand{\eeqar}{\end{eqnarray}}

\def\del{\partial}

\def\be{\begin{equation}}
\def\ee{\end{equation}}
\def\bea{\begin{eqnarray}}
\def\eea{\end{eqnarray}}
\def\ba{\begin{array}{l}}
\def\ea{\end{array}}
\def\eq#1{(\ref{#1})}
\def\tvev{{\langle {\rm T} \rangle}}
\def\dpp{{D$p$}}
\def\dppb{{$\overline{{\rm D} p}$}}
\def\Nb{{\overline N}}

\begin{document}

\begin{titlepage}

\begin{flushright}
CERN-TH/2000-090\\
TIFR/TH/00-24\\
SPhT/00-73 \\
hep-th/0005242
\end{flushright}
\vfil\vfil

\begin{center}
{\Large {\bf Supergravity Description of Non-BPS Branes}\\} 

\vspace*{15mm}
\vspace*{1mm}
Philippe Brax$^{a,c}$, Gautam Mandal$^{a,b}$ and Yaron Oz$^a$

\vspace*{1cm} 

$^a$ {\it Theory Division, CERN \\
CH-1211, Geneva, 23, Switzerland}\\

\vspace*{0.5cm}

$^b$ {\it Department of Theoretical Physics, Tata Institute of
Fundamental Research\\
Homi Bhabha Road, Mumbai 400 005, India}

\vspace*{0.5cm}

$^c${\it Service de Physique Th\'eorique, CEA \\
Gif/Yvette, F91191, France}

\vspace*{.5cm}
\end{center}

\begin{abstract}
  We construct supergravity solutions that correspond to $N$ \dpp-branes
  coinciding with $\Nb$ \dppb-branes. We study the physical properties
  of the solutions and analyse the supergravity description of tachyon
  condensation.  We construct an interpolation between the
  brane-antibrane solution and the Schwarzschild solution and discuss
  its possible application to the study of non-supersymmetric black
  holes.

\end{abstract}
\vskip 4cm

May 2000
\vfill
\hrule
e-mail: Philippe.Brax@cern.ch, Gautam.Mandal@cern.ch, Yaron.Oz@cern.ch
\end{titlepage}

\newpage

\section{ Introduction}

While a brane breaks half of the space-time supersymmetry, the
anti-brane breaks precisely the other half of the supersymmetry.
Thus, a system of a brane and anti-brane breaks together all the
space-time supersymmetry. The system is not stable, however, since the
brane and anti-brane attract each other.  This can be understood as
the appearance of a tachyon on the world-volume of the branes. It
arises from the open string stretched between the brane and the
anti-brane and it is charged under the world-volume gauge groups.  The
decay of the system can be seen by the tachyon rolling down to the
minimum of its potential \cite{senreview}.  The phenomenon of tachyon
condensation is fairly well studied by now in the open string
description \cite{sentach,numerics,hori1}.  It would be interesting to ask
how the phenomenon appears from the closed string viewpoint.  One of
the aims of this paper is to construct supergravity solutions that
correspond to N \dpp-branes coinciding with $\Nb$ \dppb-branes (anti
D-branes) and analyse the supergravity description of tachyon
condensation.

While Type IIA (Type IIB) string theory has BPS D-branes of even (odd)
dimensions, they also admit non-BPS D-branes of odd (even) dimensions.
These branes are not stable. They have been interpreted as the string
theoretical analogues of sphalerons in field theory
\cite{Harvey-Horava-Kraus}.  The families of supergravity solutions
that we will discuss contain also backgrounds that correspond to these
branes.  Stable non-BPS brane configurations are much studied too
\cite{Senpapers, BG, Witten, horava, lerda, mukhi}.  However, we will not
discuss supergravity backgrounds that correspond to these objects.

Another motivation that we have for studying brane-antibrane solutions
is to understand the relation between these solutions and the
Schwarzschild black hole solution (see, e.g.
\cite{Horowitz-Maldacena-Strominger} for an early indication of such a
connection in the context of five-dimensional black holes of Type IIB
theory), which may have possible applications in the study of
non-supersymmetric black holes.

This paper is organised as follows.  In Section 2 we describe the
supergravity solution that corresponds to N \dpp-branes coinciding
with $\Nb$ \dppb-branes and its physical properties. In Section 3 we
analyse the supergravity description of tachyon condensation.  We will
also discuss the issue of decoupling and open-closed string duality.
In section 4 we describe a general family of supergravity solutions
that includes non-Poincare-invariant world-volumes.  In particular it
contains an interpolation between the brane-antibrane solution and the
Schwarzschild solution. We discuss the possible application to the
study of non-supersymmetric black holes. Section 5 contains a
short discussion of the results.

We note that supergravity descriptions of smeared brane-antibrane
configurations have been presented in \cite{smeared}. We will discuss
in this paper the localized ones.  Unstable branes on AdS have been
analysed in \cite{Drukker}.

\section{The Supergravity Description}

In this section we will describe Type II supergravity solutions that
correspond to $N$ \dpp-branes coincident with $\Nb$ \dppb-branes and
their physical properties.

\subsection{The Supergravity Solution}

The strategy for constructing such solutions will be the
following. We know that a brane-antibrane configuration must have the full
world-volume Poincare symmetry $ISO(p,1)$\footnote{%
By contrast, a non-extremal Dp-brane
breaks $ISO(p,1)\to ISO(p)$, which is expected of a finite temperature
world-volume field theory (see Section 4). Here $I$ stands for
``inhomogeneous'', referring to the translational
symmetries.}. Furthermore, it should have
rotational symmetry $SO(9-p)$ in the $9-p$
transverse directions. For $N \neq \Nb$, the system will
also carry an appropriate RR charge.

We therefore look for the most general solution of Type II A/B
supergravity which possess the symmetry
\be
{\mathcal{S}}= ISO(p,1) \times SO(9-p) \ , 
\label{symmetry1}
\ee 
and  carries  charge under a RR field\footnote{%
Our convention for the RR field and potentials is as follows.  For
electric $p$-branes (i.e. for $p=0,1,2$), the RR field strength is $
F_{p+2} \equiv d C^{(p+1)} $.  For magnetic $p$-branes i.e. for
$p=4,5,6$, we interpret $C^{(p+1)}$ as the dual potential, and the RR
field-strength will be given by $F_{8-p} \equiv
e^{-\frac{(3-p)\phi}{2}} *\left(d C^{(p+1)}\right)$.  For 3-branes
($p=3$) the self-dual field strength is given by $ F_5 =
\frac{1}{\sqrt 2}\left( dC^{(4)} + * dC^{(4)} \right) $.}.

The most general form of the metric, dilaton and RR-field
consistent with the symmetry (\ref{symmetry1}) is 
\bea
ds^2 &=& e^{2 A(r)} dx_\mu dx^\mu + e^{2 B(r)} \left(
 dr^2  + r^2 d\Omega_{8-p}^2 \right) \ , \non
\phi &=& \phi(r) \ , \non
C^{(p+1)} &=&  e^{\Lambda(r)}\, d x^0\wedge d x^1 \wedge \ldots \wedge
 d x^p \ .
\label{metric1}
\eea

We look for  solutions of the form \eq{metric1}, of 
Type II A/B supergravity Lagrangian, 
whose relevant part is given (in the Einstein frame) by  
\be
S= \frac{1}{16\pi G_N^{10}}
\int d^{10}x \sqrt{g} \left(R- \frac{1}{2}
\del_M\phi \del^M \phi-
{1 \over 2~ n! } e^{a\,\phi} F_n^2 \right)
\label{action}
\ee
where 
$a= \frac{5-n}{2}$. The relation between the rank $n$
of the RR field strength $F_n$ and
the dimensionality $p$ of the brane has been explained in
the footnote 2.

In \eq{metric1} and in the
rest of the paper we represent ten-dimensional coordinates by $x^M,
M=0,\ldots,9$ and brane world-volume coordinates (including time) by
$x^\mu, \mu=0,1,\ldots,p$. We will
denote  the transverse coordinates by $x^i,
i=1,\ldots,9-p$  or, alternatively, by the polar coordinates $r, \theta_1,
\ldots, \theta_{8-p}\, ( r^2\equiv x^ix^i$).

The equations of motion that follow from \eq{action}
for the ansatz \eq{metric1} are  (see, e.g.,\cite{Stelle-review,Zhou})
\begin{eqnarray}
& & A'' + (p+1) (A')^2 + (7-p)\, A'B' + {8-p\over r } \, A'
 =  { 7-p\over 16 } \, S^2 \ , \non
& & B'' + (p+1)  A'B' + {p+1 \over r} \,A' +(7-p) (B')^2  
+ { 15-2p \over r}\, B' = - { 1\over 2}\, 
{p+1 \over 8}\, S^2 \ , \non
& & (p+1) A'' + (8-p)B'' + (p+1) (A')^2 + {8-p \over r}\, B' 
- (p+1) A'B'   + { 1\over 2} \, (\phi')^2 = 
{1\over 2}\, {7-p \over 8}\, S^2 \ , \non
& &\phi'' + \left( (p+1) A' + (7-p)B' + {8-p \over r}  
\right) \phi' = - { a\over 2} \, S^2  \ , \non
& & \left( {\Lambda' } \ e^{\Lambda+a\phi-(p+1) A+(7-p) B }
\, r^{8-p}
  \right)'=0 \ ,
\label{equations}
\end{eqnarray}
where
\beq
S= {\Lambda' } \, \ e^{\frac{1}{2}a\phi+\Lambda- (p+1) A} \ .
\eeq

The mathematical solution to this system of differential equations has
already  been presented in \cite{Zhou}.  The solutions
depend on three\footnote{%
One would naively predict {\sl two} parameters, corresponding to
the mass and the charge of the system, based on a suitable
generalisation of Birkhoff's theorem. However, the proofs
of such uniqueness theorems assume regular manifolds and
therefore do not apply here.}
parameters $r_0,c_1,c_2$ (we have relabelled $c_3$ of
\cite{Zhou} as $c_2$, and $k$ as $-k$) and are given by
\begin{eqnarray}
 A(r) & = &  \frac{(7-p)(3-p)c_1}{64} \, h(r) 
\nonumber \\
& & 
 - { 7-p \over 16} 
\,  \ln\left[\cosh(k \, h(r) ) 
- c_2\, \sinh( k \, h(r) ) \right],
\non
B(r) & = &    { 1 \over  7-p } \, \ln \left[f_-(r) f_+(r) \right]
+{ (p-3)(p+1)c_1\over 64 }  \, h(r) 
\nonumber \\
& &  + {p+1\over 16 } \, 
\ln\left[ \cosh( k \, h(r) ) 
- c_2\, \sinh( k \, h(r) ) \right],
\non
\phi(r) & = &    {(7-p)(p+1)c_1\over 16 }h(r)   
\non
& & + { 3-p\over 4} \, \ln\left[ \cosh(k \, h(r) ) 
- c_2\, \sinh( k \, h(r) ) \right] , 
\non
e^{\Lambda(r) } & = &  - \eta  (c_2^2-1)^{1/2} \,  {\sinh (k \, h(r) ) 
\over  \cosh( k \,  h(r) )  - c_2\, \sinh( k \, h(r) ) } \ ,
\label{solutions}
\end{eqnarray}
where 
\begin{eqnarray}
f_\pm(r) &\equiv & 1\pm \left(\frac{r_0}{r}\right)^{7-p} \ ,
\non
h(r)  &=& \ln \left[\frac{f_-(r)}{f_+(r)} \right] \ , 
\non
k & = & \pm \sqrt{{2(8-p) \over 7-p }
-\frac{(p+1)(7-p)}{16}c_1^2} \ , 
\non
\eta & = & \pm 1 \ .
\label{definitions}
\end{eqnarray}
The parameter $\eta$ describes whether we are measuring the
``brane'' charge or the ``antibrane'' charge of the system.

The parameters $(r_0,c_1,c_2)$ appear as integration constants and as
such they could be complex, describing a six-dimensional space.
However, the reality of the supergravity fields singles out three
distinct three-dimensional subspaces I, II and III, as discussed in
appendix A. For the rest of our paper, we will concentrate on the
physical properties of the solution I where the above three parameters
are all real; we will comment on II and III in appendix A. We also
note that besides the three continuous parameters $r_0, c_1$ and $
c_2$, our solution has two additional discrete parameters: sign$(k),
\eta$.

The solution is invariant under three independent
$Z_2$ transformations which act on the space of
the parameters
\be
\ba
(\mu, c_1, c_2, {\rm sign}(k), \eta)
\to 
(\mu, c_1, -c_2, -{\rm sign}(k), -\eta)
\\
(\mu, c_1, c_2, {\rm sign}(k), \eta)
\to
(-\mu, -c_1, c_2, -{\rm sign}(k), \eta)
\\
(\mu, c_1, c_2, {\rm sign}(k), \eta)
\to
(-\mu, -c_1, -c_2, {\rm sign}(k), -\eta)
\\
\mu \equiv r_0^{7-p} \ .
\ea
\label{z2s}
\ee

For convenience we will fix the above $Z_2$'s
by choosing \\
\noindent (a) 
the positive branch of the square root for $k$, namely 
\be
k= \sqrt{{2(8-p) \over 7-p }
-\frac{(p+1)(7-p)}{16}c_1^2} \ ,
\label{k-branch}
\ee
\noindent (b)
\vspace{-3ex}
\be
c_1 \ge 0 \ .
\label{c1-branch}
\ee

\vspace{10ex}

\noindent{\bf The case of the instanton ($p=-1$):}

\vspace{3ex}
The solutions mentioned above also include $p=-1$. In this
case there is no $A(r)$; the metric, dilaton 
and the RR potential are explicitly given by
\bea
ds^2 &=& \left(\frac{f_-(r)}{f_+(r)}\right)^{1/4}
 \left(
 dr^2  + r^2 d\Omega_{8-p}^2 \right) \ , \non
\phi &=&  \ln\left[ \cosh({3\over 2} \, h(r) ) 
- c_2\, \sinh({3\over 2}  \, h(r) ) \right] \ , \non
C^{(0)} &=&  e^{\Lambda(r)}= - \eta  (c_2^2-1)^{1/2} \, 
 {\sinh ({3\over 2} \, h(r) ) 
\over  \cosh({3\over 2} \,  h(r) )  - c_2\, 
\sinh({3\over 2}  \, h(r) ) } \ ,
\label{metric-minus1}
\eea
where 
\bea
f_\pm(r)& = &\left(1- (\frac{r_0}{r})^8 \right)\ , \non
h(r) &=& \ln \left[ \frac{f_-(r)}{f_+(r)} \right] \ .
\eea
An interesting point to note is that in this case the solution depends
only on two parameters $r_0, c_2$ (which are functions of mass and
charge), consistent with Birkhoff's theorem. The extra parameter
$c_1$ does not appear. According to the interpretation in the
next section it implies that there is no tachyon associated with
this solution.

The neutral case (taken as $c_2=-1$) is described by 
\bea
ds^2 &=& \left(\frac{f_-(r)}{f_+(r)}\right)^{1/4}
 \left( dr^2  + r^2 d\Omega_{8-p}^2 \right) \ , \non
\exp[\phi] &=&  \left(\frac{f_-(r)}{f_+(r)}\right)^{3/2} \ .
\label{metric-minus1-neutral}
\eea
Regarded as a IIB solution, this should be interpreted
as a $D(-1)$-$\bar D(-1)$ pair. On the other hand,
the same solution can alternatively be regarded
as a IIA solution; in that case it has a natural lift
to eleven dimensions, given by the
formula
\be
ds_{11}^2 = \exp[4\phi/3] dx_{10}^2 + \exp[-\phi/6] ds^2 \ .
\ee
It is easy to see that the eleven-dimensional metric
becomes the Euclidean Schwarzschild metric
\be
ds_{11}^2= \left(1- \frac{M}{{\tilde r}^8} \right) dx_{10}^2
+ \frac{dr^2}{1- \frac{M}{{\tilde r}^8}} + {\tilde r}^2 d\Omega_9^2
\ee
where $M = 4 r_0^8$ and $\tilde r= r\ f_+^{1/4}$. It has
been pointed out in  \cite{Harvey-Horava-Kraus} that this
metric describes the non-BPS D-instanton of type IIA \footnote{We thank Y. Lozano
for a comment on this case.}. Thus,
we see that \eq{metric-minus1-neutral}, regarded as a IIA
solution, describes the non-BPS D-instanton. This is
in keeping with our later observations about non-BPS D-branes.
The interesting point here is that in the
absence of the extra parameter $c_1$, the same neutral supergravity
solution describes both the $D(-1)\ \bar D(-1)$ pair as
well as the non-BPS $D(-1)$ brane. This is presumably
a consequence of our earlier observation that there is no
tachyon associated with this solution.

\subsection{Physical Properties}

In \cite{Zhou} the physical interpretation of the above
three-parameter solution \eq{solutions},\eq{definitions} was not
presented. We will see that it corresponds to brane-antibrane systems
along with condensates.

In a brane-antibrane system, there are two obvious physical parameters
$N$ and $\Nb$ which are the numbers of branes and antibranes
respectively. In the above supergravity solution too, there are two
obvious physical parameters: the RR charge $Q$ and the ADM mass
$M_{ADM}$, which clearly depend on $N$ and $\Nb$.  We will discuss in
Section 3 the brane interpretation of the third parameter. Before
that, however, it will be useful to discuss $Q$ and $M_{ADM}$ in
greater detail.

For convenience, we consider wrapping the spatial world-volume
directions on a torus $T^p$ of volume $V_p$ (this is always possible,
since the metric and other fields do not depend on these directions).
The RR charge $Q$, defined by an appropriate surface integral over
the sphere-at-infinity in the transverse directions
(see, e.g. \cite{Stelle-review}), is given by
\be
Q=    2\eta N_p r_0^{7-p} k \sqrt{c_2^2-1} \ ,
\label{charge-value}
\ee
where 
\be
N_p\equiv \frac{(8-p)(7-p) \omega_{8-p}V_p}{128\pi G_N^{10}} ,
\ee
and
$\omega_d= \frac{2 \pi^{(d+1)/2}}{\Gamma((d+1)/2)}$
is the volume of the unit sphere $S^d$. We have normalized
the charge $Q$ such that the BPS relation becomes $M_{BPS}=Q$. 

The ADM mass $M$ is defined, in terms of the Einstein-frame metric, by 
\cite{Maldacena-thesis, Myers-Perry} \footnote{This definition
differs from the one presented, e.g. in 
\cite{Stelle-review}, by an overall factor.}
\be
g_{00} = -1 + \frac{16\pi G_N^{10-p} M}{(8-p)\omega_{8-p}r^{7-p}}
+ {\rm higher\,order\,terms}
\label{adm-def}
\ee
where $G_N^{10-p} = G_N^{10}/ V_p$.

This gives us
\beq
M = N_p r_0^{7-p} \left[ \frac{3-p}{2} c_1 + 2 c_2 k \right].
\label{adm-value}
\eeq
Since the solution is generically non-BPS, $M$ is
different from $M_{BPS} \equiv Q$. The
mass difference is given by
\be
\Delta M\equiv M-M_{BPS}= 
N_p r_0^{7-p} \left[ \frac{3-p}{2} c_1 + 2 k
(c_2 - \sqrt{c_2^2-1}) \right] \ .
\label{delta-m}
\ee

In order to have a better understanding of the space
of solutions represented by \eq{solutions},\eq{definitions},
we now consider some  special limiting cases.

\vskip 0.5cm

{\bf The BPS case $(\bar{N}=0)$}\\

Since the BPS \dpp-brane clearly respects the symmetry
\eq{symmetry1}, it should be part of our solution
space. 

We recall \cite{HS} that the \dpp-brane solution is
given by
\beqar
ds^2 &=& f_p^{p-7\over 8} dx_\mu dx^\mu + 
f_p^{p+1\over 8}(dr^2 + r^2 d\Omega_{8-p}^2), \non
e^{\phi} &=& f_p^{\frac{3-p}{4}} \ , \non
C^{(p+1)}_{01\ldots p}&=& -\eta \frac{1}{2}(f_p^{-1}-1) \ , \non
f_p &=& 1+ \frac{\mu_0}{r^{7-p}} \ ,
\label{dp-sol}
\eeqar 
with ADM-mass $M_{Dp}$ and charge $Q$ given by
\be
M_{Dp}= Q= {\mu_0} N_p \ ,
\label{MBPS}
\ee

This solution indeed exists in a ``scaled neighbourhood'' of the point
$(r_0,c_1,c_2) = (0,c_m,\infty)$, defined by
\bea
r_0^{7-p} &=& \epsilon^{\frac{1}{2}}\bar{r}_0^{7-p} \ , \non
c_1 &=& c_m - \epsilon \frac{8 {\bar k}^2}{(p+1)(7-p) c_m}, \non
c_2 &=& \frac{\bar{c}_2}{\epsilon} \ ,
\eea
where $c_m = (\frac{32 (8-p)}{(p+1)(7-p)^2})^{1/2}$ denotes
the point where $k=0$. The second condition is
better stated as
\be
k = \epsilon^{\frac{1}{2}} \bar k \ .
\ee
The scaling is defined by the limit $\epsilon\to 0$ such that 
$\bar r_0, \bar{c}_2$ and  $\bar k $ are fixed.

It is easy to check that the solution
\eq{solutions} reduces to \eq{dp-sol} with
\be
{\mu_0} =  2 c_2 k{r}_0^{7-p}= 2\bar{c}_2\bar k \bar{r}_0^{7-p}.
\label{mu0-k-c2}
\ee

\begin{figure}[htb]
\begin{center}
\epsfxsize=4in\leavevmode\epsfbox{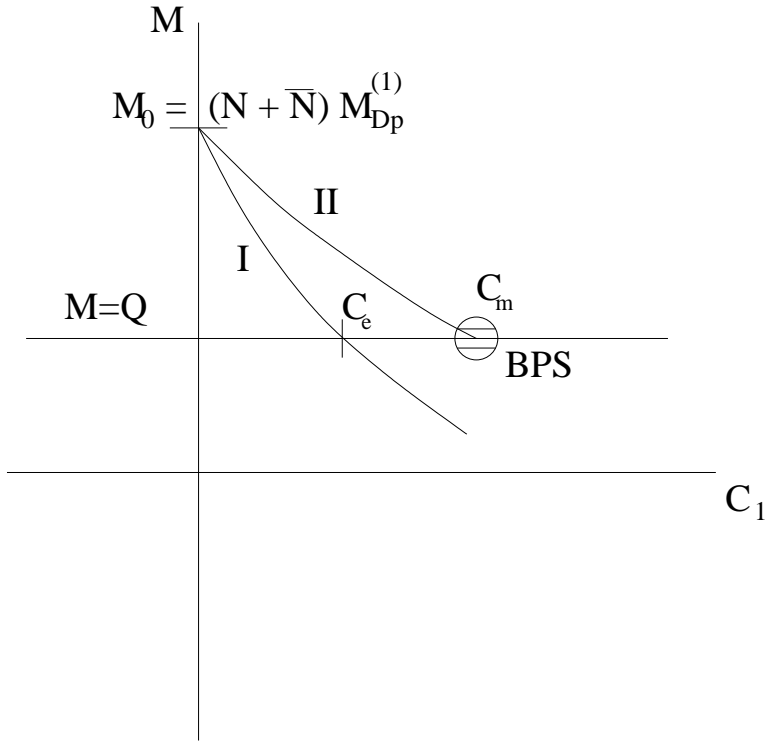}
\end{center}
\caption{The  $M,c_1$ plane for a given fixed $Q \neq 0$. The BPS
solution corresponds to the scaled neighbourhood
represented by the shaded circle. Path II represents
decay to a BPS D-brane of charge $Q$. }
\label{phasefig}
\end{figure}

It is useful to consider the three-parameter space of
solutions as parameterised by $M,Q,c_1$. Figure 
\ref{phasefig} depicts
the $M,c_1$ plane for a given fixed $Q$. The BPS
solution corresponds to the scaled neighbourhood
represented by the shaded circle. Other parts of the figure
will be explained later.

\vskip 0.5cm

{\bf The \dpp-\dppb System ($N=\Nb$)}\\

\begin{figure}[htb]
\begin{center}
\epsfxsize=4in\leavevmode\epsfbox{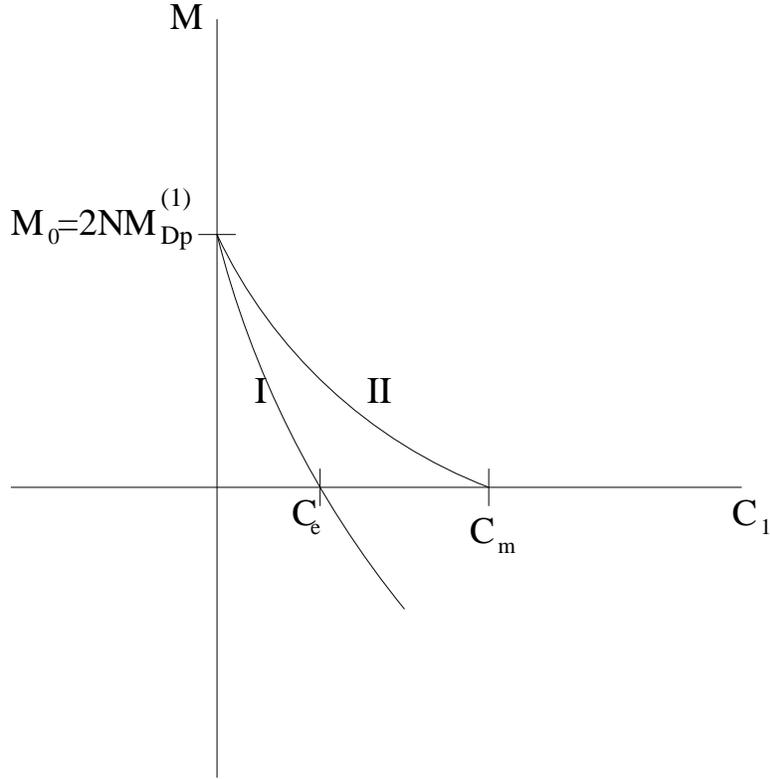}
\end{center}
\caption{The two-parameter space of
solutions for $Q=0$, as parameterised by $M,c_1$. Path II represents
decay of the brane-antibrane configuration to flat space.}
\label{neutralfig}
\end{figure}

In this case the RR charge $Q \propto (N - \Nb)$ must vanish. 
According to \eq{charge-value} this corresponds to the subspace
\be
|c_2| = 1 \ .
\label{mod-c2}
\ee
We represent this subspace in Fig \ref{neutralfig}.

Now \eq{mod-c2} implies $c_2=\pm 1$. 
As remarked in Section 3 below, the physically relevant
choice for  $p
> 3$ is $c_2=1$, while for $p <3$ it is $c_2=-1$ (for $p=3$ the two
choices are physically equivalent). To simplify the discussion we will
present the formulae in the rest of this section for $p>3$;
it is straightforward to write down the formulae in
the other cases.
  
The solution now reads 
\beqar
e^{2A} &=& \left(\frac{f_-}{f_+} \right)^{\alpha} \ ,\non
e^{2B} &=& f_-^{\beta_-}
f_+^{\beta_+} \ ,
\non
e^\phi &=& (f_-/f_+)^{\gamma} \ ,
\non
e^\Lambda &=& 0 \ ,
\label{neutral}
\eeqar 
where
\beqar
\alpha &=& (7-p)\left(\frac{(3-p)c_1+4k}{32}\right) \ , \non
\beta_{\pm} &=& \frac{2}{7-p} \mp \left(\frac{(p+1)((p-3)c_1-4k}{32} \right) 
\ , \non
\gamma &=& \frac{1}{16}\left((7-p)(p+1)c_1 - 4(3-p)k \right)  \ .
\eeqar

These represent the most general  2-parameter ($r_0,c_1$) solution
of Type II supergravity with no gauge field and SO(p,1) $\times$
SO(9-p) symmetry.

Consider for instance the case $p=6$. The solution reads
\be
\ba
e^{2A}= \left(\frac{1-r_0/r}{1+ 
r_0/r}\right)^{(4k - 
3c_1)/32} \ ,\\
e^{2B} = \left(1- r_0/r \right)^{2+7(3c_1-4k)/32}
\left(1+r_0/r\right)^{2-7(3c_1-4k)/32} \ ,\\
e^\phi= \left(\frac{1-r_0/r}{1+ r_0/r}\right)^{(7c_1+12k)/16}
\ea
\label{neutral6}
\ee
where  
$k= \sqrt{4 - 7c_1^2/16}$.

The Einstein metric has a curvature singularity at $r=r_0$. The
scalar curvature in \eq{neutral6}, e.g., goes as
\beq
{\cal R} \sim \frac{1}{(r-r_0)^{2 + \beta_-}} \ .
\label{curv}
\eeq
The physical regime is $r \geq r_0$.
In the case of a single D$p$-brane the curvature 
singularity is resolved  by the appropriate inclusion
of the brane degrees of freedom. We will
discuss this issue in our case later on.

\noindent For the specific value
\be
c_1=0 \ ,
\ee
we get 
\be
\ba
e^{2A}= \left(\frac{1-\frac{r_0}{r}}{1+ \frac{r_0}{r}}\right)^{1/4} \ ,\\
e^{2B}= (1- \frac{r_0}{r})^{1/4}(1+\frac{r_0}{r})^{15/4} \ ,\\
e^\phi= \left(\frac{1-\frac{r_0}{r}}{1+ \frac{r_0}{r}}\right)^{3/2} \ ,
\ea
\label{sen-isotropic}
\ee
which is the coincident D6-$\overline{{\rm D}6}$ solution 
\cite{gerome,Sen} in isotropic coordinates. In Fig \ref{neutralfig}, this
corresponds to the point $(M,c_1)= (M_0,0)$.

The above observation implies that for $c_1 \neq 0$ we get a
generalisation of the coincident D6-$\overline{{\rm D}6}$ solution.
We will argue in the next section that the parameter $c_1$ is related
to the ``vev'' \footnote{We actually consider generically off-shell
  values of the tachyon. The issue of why we may have supergravity
  solutions corresponding to an off-shell tachyon is discussed in Sec
  3.1.}  of (the zero momentum mode of the) the open string tachyon
arising from open strings stretched between the D6 and $\overline{{\rm
    D}6}$ (and more generally between \dpp and \dppb) branes. The Sen
solution corresponds to the particular case where the tachyon vev is
zero.

\vskip 0.5cm

\noindent{\bf Other cases of $\Delta M=0$}\\

Clearly, from \eq{delta-m} we can have 
\be M=Q 
\ee 
if we have 
\be
(3-p)/2~ c_1 + 2 k (c_2 - \sqrt{c_2^2-1})=0 \ .
\ee 
This solution (represented by $c_1=c_e$ in  Figs
\ref{phasefig},\ref{neutralfig}) is
nonsupersymmetric.  Indeed, there is
a  range of the parameters (see Figs
\ref{phasefig},\ref{neutralfig}) in which
\be M < Q \ , 
\ee 
These solutions cannot correspond to physical states
of string theory (for $Q=0$, these correspond to negative
ADM mass).

This implies that we expect additional contribution to the ADM mass
formula, coming perhaps from a better understanding of the curvature
singularity at $r=r_0$. In the case of BPS D-branes or the fundamental
string the ADM mass formula as found by the asymptotic behaviour of
the Einstein metric does represent the energy-momentum of the source
sitting at the curvature singularity. The reason our case is different
may have to do with the fact that we have a naked singularity at
$r=r_0$; a computation of the Euclidean action similar to that in
\cite{Gibbons-Hawking} indeed shows that the action receives
contribution not only from $r= \infty$, but also from $r=r_0$.

\section{Tachyon Condensation}

In the following, we will interpret the 3-parameter family of
supergravity solutions as a bound state of $N$ \dpp-branes coincident
with $\Nb$ \dppb-branes, together with a ``vev'' $v$ of the tachyon
condensate. The three parameters $r_0,c_1,c_2$ will be argued to
correspond to various combinations of the three parameters $N,\Nb,v$.

\subsection{$\tvev$ in Supergravity}

A system of $N$ \dpp-branes on top of $\Nb$ \dppb-branes has a tachyon
arising from the open string stretched between the \dpp-branes and the
\dppb-branes. The tachyon $T$  transforms in the $(N,\Nb)$
(and $T^*$ in $(\Nb,N)$)
representation of the $U(N) \times U(\Nb)$ gauge group.  Consider
first the case $N=\Nb$ (the
neutral case).  The cases that are studied most are
$N=\Nb=1$. In this case the tachyon is a complex field
($T,T^*$) that transforms
in the $(1,-1)\oplus (-1,1)$ representation of the $U(1) \times U(1)$
gauge group of the world-volume theory.  The brane system is unstable
due to the tachyon.  The tachyon has a potential $V(T)$ which is a
function of  $|T|^2$.  The \dpp-\dppb-branes
configuration is expected to decay into the closed string (Type II)
vacuum. Such a decay into the vacuum is conjectured to happen through
the process of tachyon condensation in which the zero-momentum mode of
the tachyon gets a specific vev. In particular, it is conjectured that
at the minimum of the tachyon potential, denoted by $|T|=T_0$, the
total energy of the system actually vanishes:
\beq
E= V(T_0) + 2M_{Dp} = 0 \ ,
\label{teq}
\eeq
where $M_{Dp}$ is the mass of a \dpp-brane.
Equation (\ref{teq}) has been established numerically to a very high accuracy
via open string field theory \cite{numerics}.
When $N > 1$ it was argued in \cite{Witten} that 
at the minimum of the potential all the eigenvalues
of $T_0$ are equal. In the following we will denote
$\frac{1}{N}$Tr$(TT^*)$ by $|T|^2$.

Let us ask ourselves how the above phenomenon appears from the
viewpoint of closed string theory.  We 
concentrate  on the neutral case first ($Q=0$) and on 
the charged case later.  There
are two ways of looking at the problem: \\
(a) \underbar{Real-time}: 
The physical decay process in terms of the brane (open
string) variables in which the tachyon rolls down to its minimum is
time-dependent.  The supergravity background of such a time-dependent
brane configuration is naively expected to be
time-dependent\footnote{We remark, though, that the exterior geometry
  of a pulsating spherically symmetric star is given by the static
  Schwarzschild solution, thanks to Birkhoff's theorem. It is not
  inconceivable, therefore, to have a time-dependent brane
  configuration with a static supergravity background for $r > r_0$.
  In such a case the time-dependence could presumably be discerned at
  the level of higher mass modes of the closed string (see
  \cite{GM-hair} for a similar analysis where the supergravity
  background of a BPS state does not see the ``polarisation'' of the
  state, although the higher closed
  string modes see it.)}.\\
(b) \underbar{Path-in-configuration-space}: 
One can alternatively view the decay
as a one-parameter path in the open string configuration space, which
for our purposes here is the space of values of $|T|$. Except at the
two extremities of the path ($|T|=0, T_0$), the other values of $|T|$
are not at an extremum of $V(T)$ and is therefore off-shell.  Let us
ask how such a path would appear in the closed string description.  Let
us imagine doing an experiment in which gravitons and other massless
closed string probes are scattered off the brane-antibrane system for
various values of $|T|$ as $|T|$ is varied from $0$ to $T_0$. We will
{\sl assume} here that such an experiment makes sense with off-shell
values of the tachyon \footnote{%
  Coupling on-shell bulk degrees of freedom to off-shell brane degrees
  of freedom is also familiar from AdS/CFT.}.  In principle one can
imagine coupling closed string degrees of freedom to the off-shell
tachyon through, e.g., the modified DBI action appropriate to
brane-antibrane systems. The supergravity solution away from the brane
will have the same symmetry as the brane-antibrane system, namely
\eq{symmetry1}. However, the metric and other fields must reflect the
extra parameter $|T|$. We will try to argue that the one-parameter
deformation represented by $c_1$ in our solution corresponds to this
$|T|$.

We begin by asking whether we see in the
supergravity description an analogue of the
tachyon potential.  The obvious supergravity counterpart of the
total energy $E$ (Eqn. \eq{teq}) of the brane-antibrane system 
is the ADM mass
\eq{adm-value}. For the suggested identification to be
correct we should have 
\beq 
M = V(T) + 2N M^{(1)}_{Dp}  \ .
\label{teqq}
\eeq
where by $M^{(1)}$ we mean the ADM mass for a single D$p$ brane.
The supergravity solution in question here is
the  2-parameter family (\ref{neutral})
of solutions parameterised by $(r_0,c_1)$.
Since the left hand side of \eq{teqq} is the ADM mass (\ref{adm-value}),
viz. 
\beq
M = N_p r_0^{7-p} \left[ \frac{3-p}{2} c_1 +  
\left(\frac{2(8-p)}{7-p}-\frac{(p+1)(7-p)}{16}c_1^2
\right)^{\frac{1}{2}} \right] \ ,
\label{adm}
\eeq
let us ask whether the 
the qualitative
behaviour of $M$  as a function of $c_1$ in \eq{adm}
agrees with the right hand side of \eq{teqq} 
for some appropriate identification between $c_1$
and $T$.

\noindent\underbar{Comment on branches}:
As explained in Appendix A, the dependence 
of the ADM mass on $c_1$ depends on the specific
branch of the solution. In the following we
will find that it is for the branch $I_{++}$
for $p > 3$ (and $I_{--}$ for $p < 3$)\footnote{For
$p=3$ and $Q=0$ $I_{++}$ and $I_{--}$ are physically indistinguishable.}
which lends to a tachyon interpretation. 
Later on we will briefly comment on the possible interpretation of the
other branches.

\begin{figure}[htb]
\begin{center}
\epsfxsize=4in\leavevmode\epsfbox{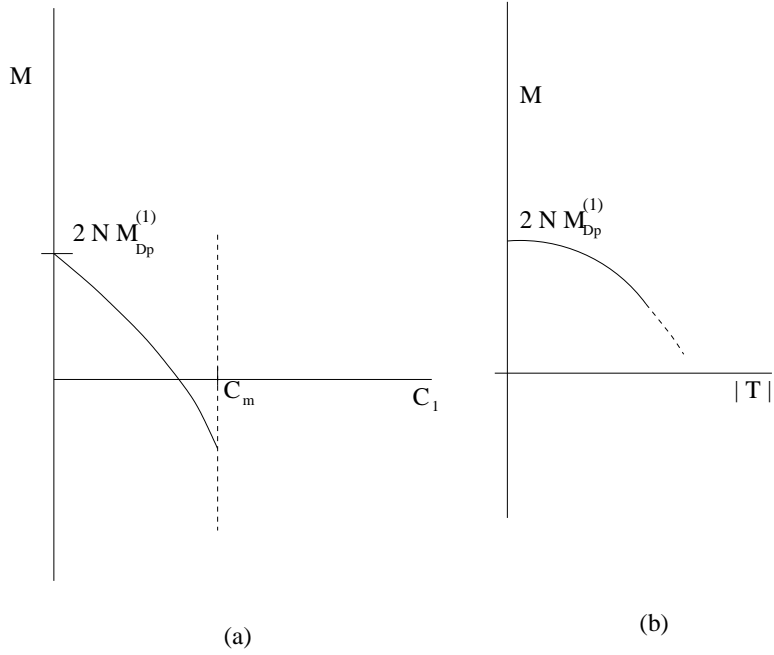}
\end{center}
\caption{ADM Mass (for a
fixed $r_0>0$) as a function of (a) $c_1$ and (b) $|T|$}
\label{admfig}
\end{figure}

Once we choose the appropriate branch of the supergravity
solution, the qualitative behaviour of $M$ as a function
of $c_1$ (at a fixed $r_0$) is given by Fig \ref{admfig}.

Consider first the case $p=6$.  When $c_1=0$ we have the coincident
D6-D$\bar 6$ solution \cite{gerome,Sen}.  The ADM mass \eq{adm} for
$p=6,c_1=0$ is $M= 4 N_p r_0$. 
We will argue in Sec. 3.2 that this mass coincides with
\be 
M= 2 N M_{Dp}^{(1)} \ .
\ee 
This implies that $V(T) = 0$ at $c_1=0$; since
 the tachyon potential vanishes only
at $T=0$ \cite{senpot}, we conclude that 
\be
T=0 ~~~~~~{\rm at}~~~~~~ c_1=0 \ .
\label{t0-at-c0}
\ee
As we will see, the last equation is valid for all $p$.
This will imply that the subspace of our
three-parameter solution defined by $c_1=0$
represents \dpp-\dppb branes with zero value of the
tachyon $|T|$, that is, brane-antibrane configurations
which sit at the maximum of the tachyon potential.

Let us now consider small deformations away from $c_1=0$.
Since $V(T)$ is known to be a function only of $|T|^2$, we
expect the ADM mass, and hence $c_1$, to be a function
of $|T|^2$ too. For small deformations, we can write
\be
c_1 = a |T|^2 + b |T|^4 + \ldots
\label{c1-vs-t}
\ee
Clearly  $a>0$. It is easy to see that the behaviour of the ADM mass $M$
as a function of $|T|$ (Fig. \ref{admfig}(b)) qualitatively
matches the behaviour of $V(|T|)$ near $T=0$.

\vspace{5ex}

\noindent{\bf Tachyon condensation}

In Fig \ref{admfig}(b) we have not plotted the ADM mass in the whole
range of $|T|$ because \eq{c1-vs-t} is valid only near $T=0$. The
question then is whether our solution can describe the full
double-well potential $V(T)$. In other words, can we describe the
process of tachyon condensation all the way to the vacuum?

In Fig \ref{neutralfig}, vacuum is represented by
any point in the line $M=0$. Any path connecting
the point $(M_0,c_1=0)$ to this line (e.g. path I
or path II) therefore in principle represents 
a family of supergravity solutions corresponding to
a flow of $|T|$ from $|T|=0$ to $|T|= T_0$. 

To know what the actual path is, we need to have a more
precise knowledge of mapping (more
detailed than \eq{c1-vs-t}) between the open
string variables  $(N, |T|)$ to
the supergravity variables $ (r_0, c_1)$. Assuming
that such maps exist and are smooth and
invertible, the generic form will be
\be
\ba
r_0= \tilde f_1(N,|T|^2), c_1= \tilde f_2(N, |T|^2)
\\
N= \tilde g_1(r_0,c_1), |T|= \tilde g_2(r_0,c_1)
\ea
\label{map1}
\ee
These can alternatively be stated as
a map $(N, |T|) \to (M, c_1)$:
\be
\ba
M= f_1(N,|T|^2), c_1= f_2(N, |T|^2)
\\
N= g_1(M,c_1), |T|= g_2(M,c_1)
\ea
\label{map2}
\ee
Of course \eq{map1},\eq{map2} should be consistent with  \eq{c1-vs-t}
near $T=0$ (we need to consider the coefficients $a,b,\ldots$
to be functions of $r_0$ or $N$).   

The path I in Fig \ref{neutralfig} corresponds, in terms
of \eq{map1}, to $r_0= \tilde f_1(N)$ and $c_1=
\tilde f_2(|T|^2)$. This path corresponds to the plot
Fig \ref{admfig}(a) of $M$ as a function of $c_1$
{\em at fixed $r_0$}. It has the unphysical feature
that it does not stop at $M=0$ and goes down to the
domain of $M<0$. 

Path II in Fig \ref{neutralfig} requires the functions $\tilde
f_{1,2}$ (or the functions $f_{1,2}$) to be necessarily a function of
two variables. In other words, the flow of $|T|$ from $0$ to $T_0$
should mean here that both $r_0$ and $c_1$ should change appropriately
to take the solution to the point $(M,c_1) = (0, c_m)$.  The nice
feature of this path is that it automatically {\sl ends} at the flat
space solution, since $c_1$ cannot go beyond $c_m$ (actually there is
another branch of solution (Branch II, appendix A) for $c_1>c_m$, but
it can be shown that the ADM mass increases for $c_1>c_m$).

In the absence of a decoupling limit (as we will discuss in Section 3.3)
it may not be possible to determine the exact functions mentioned in
\eq{map1} or \eq{map2} and therefore to know any more about the nature
of $V(T)$ than what we have already presented here. In any case, if an
analysis of brane degrees of freedom is expected to remove the $M<0$
region, presumably the formulae for the mass will change.

In summary, we see that a path exists (path II in Fig
\ref{neutralfig}) in our space of solutions which describes the flow
of $|T|$ from $0$ to $T_0$ and the behaviour of the ADM mass $M$ along
this path matches the qualitative features of $V(T)$.

\vspace{5ex}

\noindent{\bf The other branches}

In the above we have discussed only the branch $I_{++}$ (see Appendix
A for notation) for $p \ge 3$ and $I_{--}$ for $p < 3$. It is easy to
see that the behaviour of the branches $I_{-+}, I_{+-}$ are outright
unphysical.  This leaves $I_{--}$ for $p \ge 3$ and $I_{++}$ for $p <
3$. In this branch (except for $p=3$) for small deformations of $c_1$
away from zero, $M$ initially rises beyond the combined rest mass of
the brane-antibrane system and then falls again.  This seems puzzling
since equation (\ref{teqq}) does not allow such an increase in the
energy of the system.  We should recall however that when the vev of
the tachyon field is zero the world-volume gauge group is not broken.
That means that we are allowed to have other condensates such as a
gluon condensate.  This can increase the energy of the system. An
estimate of such an increase can be obtained from the modified DBI
action \cite{Berg} 
\be S= -T_p \int d^{p+1}\sigma
 e^{-\phi} V(T) \sqrt{{\rm det} \left[G_{ij} +
  2\pi\alpha' \left(F_{ij} + \partial_i T \partial_j T
\right)\right]} \ .  
\label{DBI}
\ee 
The interpretation of the $c_1$ deformation (for $p \not=
3$) in these
branches could therefore be in terms of a gluon 
condensate. However,it remains a mystery in that case
why (a) there is no such phenomenon for $p=3$ (since
the branches $I_{++}$ and  $I_{--}$ appear to be identical),
and (b) why the ADM mass starts to decrease after a while.

\vspace{5ex}

\noindent{\bf Non-BPS D-branes}
 
Since we are only discussing the tachyon condensate in terms of a real
quantity $|T|$ we are left with the possibility that our supergravity
solution may represent a real tachyon as well.  Recall that a real
tachyon characterizes the non-BPS Dp branes, i.e.  $p$ odd for IIA and
$p$ even for IIB, which are obtained from the \dpp-\dppb-brane system
by a $(-1)^{F_L}$ projection.  So the natural question arises: which
brane system does the supergravity solution describe.  It is plausible
that in the neutral case the solution describes both.  In both cases
the background has no RR charge, and one expects the full
$SO(p,1)\times SO(9-p)$ symmetry.  The solution (\ref{neutral}) is the
most general one that satisfies these conditions.  The question is
whether the ADM mass of a non-BPS brane (with or without tachyon)
occurs in these solutions.  We recall that the tension of non-BPS Dp
branes (for $N=1$) is related to the tension of the \dpp-\dppb-brane
system by $M_{{\rm non}\hbox{-}{\rm BPS}} = \frac{1}{\sqrt{2}}
M_{Dp\hbox{-}\overline{Dp}}$, reflecting a bound system. For $N>1$
too, the tension of the non-BPS Dp brane system $M_{{\rm
    non}\hbox{-}{\rm BPS}}^{(N)}$ should be less than that of the
combined rest mass $2NM_{Dp}^{(1)}$ of the brane-antibrane system.
Since the values of ADM mass discussed in the context of (\ref{teqq})
range all the way from $2NM_{Dp}^{(1)}$ to $0$, we see that in a
suitable range of parameters the solution \eq{neutral} does have ADM
masses that can be fitted to $M= M_{{\rm non}\hbox{-}{\rm BPS}}^{(N)}
+ \tilde V(T)$ where $\tilde V(T)$ is the potential for the real
tachyon in this case.

\vspace{5ex}

\noindent{\bf The charged case: $Q\ne 0$}

In this case we expect the relation
\be
M = (N+\Nb) M^{(1)}_{Dp} + V(T) \ .
\ee
where $M^{(1)}_{Dp}$ denotes the ADM mass for a single
D$p$ brane.
The analysis of the binding energy in the next section once again
suggests that $c_1=0$ corresponds to the point where the tachyon
potential vanishes, which we expect to be for vanishing tachyon field.
The discussion of tachyon condensation is similar to the neutral case.
Again path II in Fig \ref{phasefig} is more physical than path I
because the former ends at the BPS point and does not go to the region
$M<Q$.  The qualitative behaviour of $M$ along this path again matches
the qualitative features of a tachyon potential which has a local
maximum at $|T|=0$ and a minimum at $|T|=T_0$ where we denote
$\frac{1}{N}$Tr$(TT^*)$ by $|T|^2$ (we assume that 
all the eigenvalues of $TT^*$ are the same, namely
$T_0^2$, at the minimum).  We expect
that at the minimum $V(T) = (|N-\Nb| - (N+\Nb)) M^{(1)}_{Dp}$.

\subsection{Dp-brane probes and Binding energy}

In the last section we mentioned that $V(T)=0$ corresponds to $c_1=0$.
We derive this in the present section.

We will consider the general 3-parameter solution parametrized by
$(r_0,c_1,c_2)$.  Let us define the binding energy of the
\dpp-\dppb-branes solution to be
\be
E_B= (N + \Nb)M_{Dp}^{(1)} - M \ ,
\label{binding-def}
\ee
where $M$ is given by \eq{adm-value} and $M_{Dp}^{(1)}$ represents the
rest mass of a single \dpp-brane (or \dppb-brane), given by
(\ref{MBPS}) with the scale parameter $\mu_0 = \mu_0^{(1)}$, which
depends on $g_{str}$ and $p$, the dimensionality of the brane.

In view of the equation \eq{teqq},
\be
E_B = - V(T) \ .
\label{eb-vs-vt}
\ee

A straightforward comparison between $(N + \Nb)M_{Dp}$ and $M$ of
\eq{adm-value} is hampered by the fact that we do not know a priori
the relation between the two parameters $r_0$ and $\mu_0$ that
characterise the respective solutions \eq{solutions} and \eq{dp-sol}.
We will find this relation by the following strategy.

We consider the static force between a \dpp-\dppb-branes system and a
\dpp-brane probe (respectively a $\overline{Dp}$-brane probe) at a
distance $r$. This can be computed in two ways:

(a) From supergravity:
\be
S_{\rm probe}= -\frac{1}{g_s l_s^{p+1}}
\int d^{p+1}\sigma (e^{-\phi}\sqrt{\hat G}\pm C_{p+1})
\label{s-probe}
\ee
where $G_{MN}= e^{\phi/2} g_{MN}$ represents the string
frame metric corresponding to the solution \eq{solutions}
and $\hat G$ is its pull-back to  the world-volume.
For a \dpp  (resp. \dppb) probe, we use the
upper (resp. lower) sign.
 
Subtracting the flat space DBI part, and keeping
only the leading term in the $1/r$ expansion we get 
\be
S_{\rm probe}= 2k \frac{V_p}{g_s l_s^{p+1}}
(\frac{r_0}{r})^{7-p}(c_2 \mp \sqrt{c_2^2-1}) \ .
\label{s-probe-value}
\ee

(b) By a  string theory computation: 
\be
\langle Dp  D\bar p| \exp(-\beta H) | Dp \rangle
\label{boundary-dpdp}
\ee
where the states are regarded 
as  boundary states constructed out
of closed-string oscillators.
(We consider here the case of the \dpp-probe first.)
At weak coupling and for $\tvev=0$, the
boundary state on the left is given by
\be
\langle Dp  D\bar p|= \langle Dp | \otimes \langle D\bar p| \ .
\label{tensor-product}
\ee 
We will assume that \eq{tensor-product} can be used for
computation of the leading term in the $1/r$ expansion for large
distances $r$, when $\tvev=0$ (see \cite{DiVecchia,Panda}
for earlier work on connection between
boundary states and classical solutions).  Since the
static force between two Dp-branes vanishes, the computation (b) then
reduces, at $\tvev=0$, to
\be
\langle  D\bar p| \exp(-\beta H) | Dp \rangle \ .
\label{dpbar-dp}
\ee
This latter can be computed at large distances from supergravity, by
the DBI action of a \dpp-brane probe in the background of a \dppb-
brane:
\be
S'_{\rm probe}\equiv - \frac{1}{g_s l_s^{p+1}}
\int d^{p+1}\sigma [e^{-\phi}\sqrt{\hat G}
+ C^{(p+1)}] \ ,
\label{sp-probe}
\ee
where the metric, dilaton and the RR potential are
now obtained from \eq{dp-sol}, with $ \mu_0
=\Nb\mu_0^{(1)}$. 
We get, again after subtracting the flat space DBI part, and 
keeping only the leading term in the $1/r$ expansion,
\be
S'_{\rm probe}= \frac{V_p}{g_s l_s^{p+1}}
\left(2 \frac{\Nb \mu_0^{(1)}}{r^{7-p}}\right) \ .
\label{sp-probe-value}
\ee
This result holds for the \dpp-probe. For the
\dppb-probe we need to replace $\Nb
\to N$ in the above expression.

Matching \eq{s-probe-value}
and \eq{sp-probe-value} leads to
\begin{eqnarray}
N \mu_0^{(1)}&=& k r_0^{7-p}(c_2+ \sqrt{c_2^2 -1}) \ ,\nonumber \\
\Nb \mu_0^{(1)} &=& k r_0^{7-p}(c_2 -\sqrt{c_2^2 -1})\ .\nonumber \\
\label{mu-vs-r0} 
\end{eqnarray}
from this we deduce that
\be
Q= N_p \mu_0^{(1)} (N -\Nb)
\label{charge-vs-mu}
\ee
and 
\be \mu_0^{(1)} (N + \Nb)
=2kr_0^{7-p}c_2\ .
\label{total-mu}
\ee

Using \eq{adm-value}, (\ref{MBPS}) and  \eq{mu-vs-r0}
we can  find the zero of the binding energy
\eq{binding-def} of the \dpp-\dppb bound state. We get 
\be
E_B=0= N_p r_0^{7-p} \left[\frac{3-p}{2} c_1  \right] \ .
\label{binding-value}
\ee
Clearly $E_B$ vanishes at $ c_1=0$\footnote{
  The case $p=3$ is subtle and we extrapolated the result to this
  value of $p$ from the other values.  
  An alternative way would presumably be to use some other probe.}. 
In view of the identification \eq{eb-vs-vt}, this implies that
\be
c_1=0 \Rightarrow V(T)=0 \ ,
\ee
as promised in the last section. Note that\\
\noindent (a) If we put $c_1=0$
in \eq{adm-value} we indeed get $M= M_{Dp} + M_{\overline{Dp}}$,
consistent with the vanishing of the binding energy,\\
\noindent (b) Equations \eq{charge-vs-mu} and \eq{total-mu}
give us essentially $N- \Nb$ and $N + \Nb$ in terms
of the supergravity parameters in the subspace $c_1=0$, \\
\noindent (c) The expression for the total 
mass \eq{total-mu} matches
exactly with the BPS mass \eq{mu0-k-c2} (recall that
at the BPS point $\Nb =0 $.

\subsection{Open-closed String Duality}

In the spirit of the AdS/CFT correspondence (for a review see
\cite{us}), it is natural to ask whether we can apply a decoupling
limit \cite{Mal} of the brane modes from the bulk modes to the
supergravity description of the \dpp-\dppb-branes system.  Typically
for Dp-branes this is a low energy limit with the resulting background
being the near-horizon metric. In the present case, the closest analogue
of the near horizon metric is some suitably scaled neighbourhood of
$r= r_0$.  However, it is easy to see that for the neutral solution
\eq{neutral} there is no such region which by itself is a solution of
the supergravity field equations.  Also, we cannot find an appropriate
rescaling that keeps a metric finite in $l_s$ units as $l_s
\rightarrow 0$.  This means that the interactions between the open and
closed strings remain relevant.

\begin{figure}[htb]
\begin{center}
\epsfxsize=4in\leavevmode\epsfbox{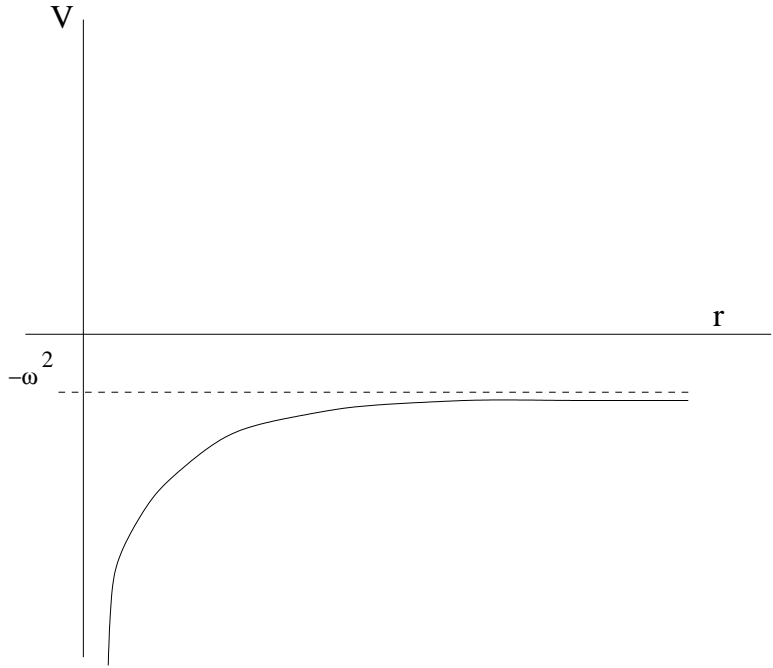}
\end{center}
\caption{The scattering potential $V(r)$
for gravitons on \dpp-\dppb-branes.}
\label{Dppfig}
\end{figure}

Another manifestation of this issue is the form of the potential
$V(r)$ for a graviton scattered on the \dpp-\dppb-branes.  The
potential is depicted in figure \ref{Dppfig}.  Near $r=r_0$ it goes
like $\frac{-1}{(r-r_0)^2}$ while at infinity it approaches
$-\omega^2$ where $\omega$ is the frequency of the scattered
graviton. The potential poses no barrier for the gravitons sent from
infinity to reach the $r=r_0$ and their absorption cross section does
not vanish \footnote{For a similar but detailed analysis see
  \cite{AIO}.}.  The absence of a decoupling of the closed strings
from the open strings prevents us from making a precise correspondence
between the field theory on the \dpp-\dppb-branes world-volume and the
supergravity background.  This suggests that there is also a
limitation on the quantitative understanding of the tachyon
condensation process by using only the open string description.

The singularity of the supergravity solution at $r=r_0$ is time-like.
Having such a singularity of the classical geometry which we can reach
at a finite proper time, there is the natural question whether it is
resolved quantum mechanically. One criterion \cite{HM} is the
existence of a self-adjoint Laplacian. This can still be the case even
if the metric is geodesically incomplete. The requirement is the
existence of a non-normalizable solution of the wave equation. This
criterion is satisfied by our geometry.
To see that we consider the Laplace equation in the form
\begin{equation}
\frac{\partial^2 \phi}{\partial t^2}=-A\phi \ .
\end{equation}
The equation $A\phi=\lambda \phi$ takes the form
\begin{equation}
\rho^{\beta}\partial_{\rho}(\rho\partial_{\rho}\phi)=\lambda\phi \ ,
\end{equation}
where $\beta=2p-15 +2(7-p)(\frac{3-p}{8}c_1-\frac{k}{2})$ and $\rho = r-r_0$.
Defining $z=\sqrt \lambda \rho^{(1-\beta)/2}$ we get 
\begin{equation}
\phi''+\frac{\phi'}{z}+\frac{4}{(1-\beta)^2}\phi=0 \ .
\end{equation}
This has Bessel function solutions  behaving like
$1$ and $\ln\rho$.
The norm of the latter $\int d\rho \rho \rho^{-2}$ diverges.

\section{The Four-Parameter Solution}

In this section we will briefly describe the most general $p$-brane
solution of Type II string theory in which we relax the requirement of
Poincare invariance in the $(p+1)$ dimensional world-volume. In
other words, we ask ourselves about the most general solution
which respects the symmetry
\be
{\mathcal S}' = SO(p) \times SO(9-p) \ ,
\label{symmetry2}
\ee
Clearly the previous 3-parameter solution already respects
this symmetry and hence should be part of this most
general family of solutions. The modified ansatz for 
the Einstein metric is 
\be
ds^2 =  \exp(2 A(r)) (-f(r)dt^2 + dx_m dx^m) + \exp(2 B(r)) \left(
 dr^2  + r^2 d\Omega_{8-p}^2 \right) \ ,
\label{metric2}
\ee 
where we split the world-volume index $\mu$ as $0, m=1,\ldots,p$.
The ansatz for the dilaton and the gauge potential remain the same as
in \eq{metric1}. 

\begin{figure}[htb]
\begin{center}
\epsfxsize=4in\leavevmode\epsfbox{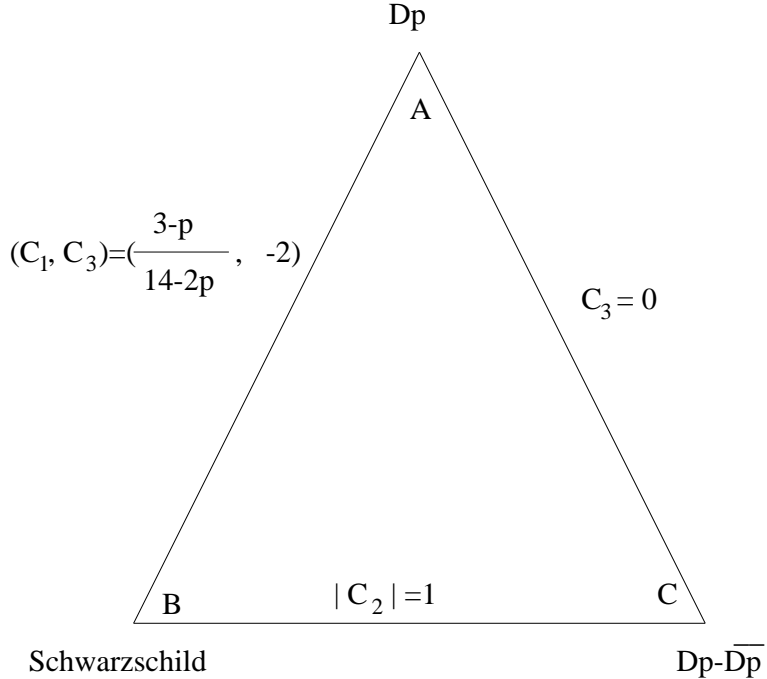}
\end{center}
\caption{The most general spherically symmetric solution
of Type II theories}
\label{generalfig}
\end{figure}

The equations of motion for this ansatz have been written down in
appendix B. Once again the mathematical solution of the differential
equations has been worked out in \cite{Zhou}. We write the explicit
solution in Appendix B for completeness and discuss here some salient
physical features (see Fig \ref{generalfig}) 

$\bullet$ The general solution has 4 independent parameters $(r_0,
c_1,c_2,c_3)$. The Poincare-invariant 3-parameter subspace discussed
in the previous sections corresponds to $c_3=0$. In Fig
\ref{generalfig} this is schematically represented by the arm AC of
the triangle ABC.  $c_3\ne 0$ breaks world-volume Poincare invariance.

$\bullet$ The two-parameter subspace
$(c_1,c_3)=(\frac{3-p}{2(7-p)},-2)$ corresponds to the black $p$-brane
solutions of \cite{HS}. This has already been identified in
\cite{Zhou}.  In Fig \ref{generalfig}, this is represented by the arm
AB of the triangle. Recall that the black $p$-branes are parameterised
by their charge and mass (equivalently $r_+, r_-$, the outer and inner
horizons). Note that the BPS D-brane can be reached as a limit
along the arm BA, like it can be reached along CA,
although the $c_3$ values characterising
these two arms are different. It is likely that there are
continuous families of solutions between BA and CA
(corresponding to different $c_3$ values) which can
reach the BPS solution under a limiting procedure. 

$\bullet$ The three-parameter subspace defined by $|c_2|=1$ describes
the most general spherically symmetric solution with no gauge fields.
This is represented by the arm BC of the triangle. It is well-known
that the neutral limit of the black $p$-brane (point B) corresponds to
the Schwarzschild black hole in $10-p$ dimensions ($\times T^p$, assuming a
wrapped $p$-brane). On the other hand, as discussed
at great length in this paper, the neutral limit of
the arm AC corresponds to the coincident brane-antibrane solutions.
The arm BC therefore provides interpolating solutions
which connect the brane-antibrane solution to the Schwarzschild
solution.

It is clear that there is a rather rich phase structure in Fig
\ref{generalfig}. Parts of this diagram have obvious decoupling limits
and dual field theory descriptions. It would be interesting to chart out
these parts completely \cite{work-in-progress}.  

Interpolations similar to the arm BC are of paramount importance to
the study of the D1/D5 system and the five-dimensional black hole
\cite{GM-review}. It has been found that CFT descriptions
seem to work in some contexts for non-rotating BTZ black holes
which are the analogues of Schwarzschild black holes in
AdS$_3$. An interpolation of such a solution to a
brane-antibrane solution of the D1/D5 system would shed
light on both brane-antibrane dynamics and nonsupersymmetric
black holes.

It has been pointed out by \cite{Gibbons-Maeda} that the equations of
motion of the above system are identical to those of a Toda molecule.
It is tempting to construct a ``mini-superspace'' kind of model for
this space based on Toda dynamics.

\section{Discussion}

In this paper we constructed localised supergravity solutions
corresponding to bound states of $N$ \dpp-branes coinciding with $\Nb$
\dppb-branes for $p=0,1,\ldots,6$ (and 
non-BPS D-branes of odd (even) dimensions of 
Type IIA (Type IIB) string theory)%
\footnote{The case $p=-1$ has been mentioned separately
in Section 2}.
We constructed these by looking for
the most general solution of Type II A/B supergravity
(in the presence of a single RR gauge field) which respect
world-volume Poincare invariance and rotational invariance in the
transverse directions. Contrary to the naive expectation that the
solution should have only two parameters corresponding to the charge
and the mass, we found that the most general solution has one extra
parameter. We found that in the physically relevant branch there are
two special values of the extra parameter  at which the
ADM mass respectively coincides with (a) the combined rest mass of the branes and
antibranes, and (b)  the mass of the BPS configuration of $N- \Nb$ branes
\footnote{for $N>\Nb$; for $N<\Nb$ these will be $\Nb- N$
  antibranes.}.  In the case $N=\Nb$ (zero RR charge) the point (b)
represents flat space.  The case $N=\Nb$ is extensively studied from
the point of view of open strings living on the brane-antibrane
system, and we recognised the solutions (a) and (b) as the supergravity
background corresponding to the maximum and the minimum of the tachyon
potential. This lead us to interpret the extra parameter in our
solution as the supergravity manifestation of an expectation value of
the tachyon.  We matched the qualitative behaviour of the ADM mass as a
function of this extra parameter with the behaviour of  the
tachyon potential $V(T)$.

We noticed the absence of a decoupling of the bulk closed strings from
the brane-antibrane open strings.  This means that the interactions
between the open and closed strings remain relevant and suggets that
there is also a limitation on the quantitative understanding of the
tachyon condensation process by using only the open string
description.

We briefly discussed a more general (four-parameter) space of
solutions in which we assume only rotational invariance in the spatial
directions on the world-volume. This space includes brane-antibrane
pairs, BPS D-branes, the black $p$-branes of \cite{HS} and
Schwarzschild black holes. The detailed understanding of this
four-parameter space in terms of brane variables is an
outstanding problem.

\vspace{5ex} {\bf Acknowledgement}:
We would like to thank A. Kumar and P. Townsend for  discussions.

\newpage

\appendix{Real Sections of the Supergravity Solution}

As remarked in the text, the three parameters $(r_0,c_1,c_2)$
characterising the supergravity solution \eq{solutions},\eq{definitions}
appear as integration
constants in the solution of differential equations
and as such could be complex. However, this
would generically make the metric, dilaton and gauge field also
complex.  We find that there are three distinct 3-dimensional domains of
$(r_0,c_1,c_2)$, described below as Branches I, II
and III, where the supergravity fields remain
real.

\noindent{\bf  Branch I}:
\be
\ba
c_1 \in (0, c_m), \quad c_m = \sqrt{8-p\over 8(p+1)(7-p)}
\\
c_2 \in (-\infty,1) \cup (1,\infty)
\\
\mu \equiv r_0^{7-p}  \in R
\\
\eta=\pm 1
\ea
\ee
We will assume in this section that  we have already fixed the $Z_2$
symmetries \eq{z2s} of the solution by implementing
\eq{k-branch},\eq{c1-branch}. For Branch I, 
the remaining choices of signs are best discussed by
thinking of four sub-branches,
depending on whether the signs of $(c_2,\mu \equiv
r_0^{7-p})$ are
$++, +-, -+$ and $--$ respectively. We denote these
as $I_{++}, I_{+-}, I_{-+}, I_{--}$ respectively (each
of these will also contain $\eta= \pm$).
The formulae for the ADM mass and charge
for Branch I is given by  \eq{adm-value},\eq{charge-value}.
Explicitly
\be
\ba
M=N_p r_0^{7-p} 
\left[ \frac{3-p}{2} c_1 + 2 c_2 
\sqrt{ \frac{2(8-p)}{7-p} -
  \frac{(p+1)(7-p)}{16}c_1^2}
\right]
\\
Q=2 \eta N_p r_0^{7-p} \sqrt{ \frac{2(8-p)}{7-p} -
  \frac{(p+1)(7-p)}{16}c_1^2} \sqrt{c_2^2-1} \ ,
\ea
\ee
The behaviour of these functions depends on
the signs of $c_2$ and $\mu$.
We find that it is the branch $I_{++}$
for $p=3,4,5,6$ which lends to a tachyon interpretation
(Section 3). For $p=0,1,2,3$ it is $I_{--}$.

\vspace{5ex}

\noindent{\bf Branch II}

\be 
\ba 
c_1 \in (c_m,\infty) \Rightarrow k = -i
\tilde k, \tilde k= \sqrt{ -\frac{2(8-p)}{7-p} +
  \frac{(p+1)(7-p)}{16}c_1^2}
\\
c_2 = i \tilde c_2, \tilde c_2 \in R
\\
\mu \equiv r_0^{7-p} \in R 
\\
\eta= \pm 1
\ea 
\ee

The mass and charge for this branch read
\be
\ba
M=N_p r_0^{7-p} \left[ \frac{3-p}{2} c_1 + 2 c_2 
\sqrt{ - \frac{2(8-p)}{7-p} +
  \frac{(p+1)(7-p)}{16}c_1^2}
\right]
\\
Q=2 \eta N_p r_0^{7-p} \sqrt{ \frac{2(8-p)}{7-p} -
  \frac{(p+1)(7-p)}{16}c_1^2} \sqrt{(\tilde c_2)^2+ 1} \ ,
\ea
\ee

\vspace{5 ex}

\noindent{\bf Branch III}

\be 
\ba 
c_1 = i \tilde c_1, \tilde c_1 \in R^{+}
\\
c_2 = i \tilde c_2, \tilde c_2 \in R
\\
\mu \equiv r_0^{7-p} = -i \tilde \mu, \tilde \mu \in R 
\\
\eta= \pm 1
\ea 
\ee

The mass and charge for this branch read
\be
\ba
M=N_p r_0^{7-p} \left[ \frac{3-p}{2} c_1 + 2 c_2 
\sqrt{ \frac{2(8-p)}{7-p} +
  \frac{(p+1)(7-p)}{16}(\tilde c_1)^2}
\right]
\\
Q=2\eta N_p r_0^{7-p} \sqrt{ \frac{2(8-p)}{7-p} +
  \frac{(p+1)(7-p)}{16}(\tilde c_1)^2} \sqrt{(\tilde c_2)^2+ 1} \ ,
\ea
\ee

\section{Details of the 4-parameter solution}

The equations of motion that follow from \eq{action}
for the ansatz \eq{metric2} are  
\begin{eqnarray}
& & A'' + (p+1) (A')^2 + (7-p)\, A'B' + {8-p\over r } \, A'
+\frac{1}{2}(\ln f)'\,A' =  { 7-p\over 16 } \, S^2 \ , \non
& & A'' + (p+1) (A')^2 + (7-p)\, A'B' + {8-p\over r } \, A'
+\frac{1}{2}(\ln f)'' + \non
& &\frac{1}{2}(\ln f)'\left((d+1) A'
+ \frac{1}{2}(\ln f)' + (7-p) B' +
\frac{8-p}{r} \right)  =  { 7-p\over 16 } \, S^2 \ , \non
& & B'' + (p+1)  A'B' + {p+1 \over r} \,A' +(7-p) (B')^2  
+  \frac{1}{2}(\ln f)' \left(B' + \frac{1}{r}\right) + \non
& & { 15-2p \over r}\, B' = - { 1\over 2}\, 
{p+1 \over 8}\, S^2 \ , \non
& & d A'' + (8-p)B'' + (p+1) (A')^2 + {8-p \over r}\, B' 
- (p+1) A'B'   + \non
& & \frac{1}{2}(\ln f)'' + \frac{1}{4} \left( (\ln f)'\right)^2 +
 { 1\over 2} \, (\phi')^2 = 
{1\over 2}\, {7-p \over 8}\, S^2 \ , \non
& &\phi'' + \left( (p+1) A' + (7-p)B' + {8-p \over r}  
+\frac{1}{2}(\ln f)' \right) \phi' = - { a\over 2} \, S^2  \ , 
\non
& & \left(\frac{\Lambda' }{f^{\frac{1}{2}}}
  \ e^{\Lambda+a\phi-(p+1) A+(7-p) B }
\, r^{8-p}
  \right)'=0 \ ,
\label{equations1}
\end{eqnarray}
where
\beq
S= \frac{\Lambda' }{f^{\frac{1}{2}}} \, 
\ e^{\frac{1}{2}a\phi+\Lambda- d A} \ .
\eeq

The solutions \cite{Zhou} 
depend on four parameters $r_0,c_1,c_2,c_3$ (we have interchanged
the labels 
$c_2,c_3$ for convenience, compared to \cite{Zhou}),  
and are given by
\begin{eqnarray}
f(r) &=& e^{- c_3 h(r)}, \non
 A(r) & = &  \frac{(7-p)}{32}\left( \frac{3-p}{2}c_1
+ (1 + \frac{(3-p)^2}{8(7-p)}) c_3 \right) \, h(r) 
\nonumber \\
& & 
 - { 7-p \over 16} 
\,  \ln\left[\cosh(k \, h(r) ) 
- c_2\, \sinh( k \, h(r) ) \right],
\non
B(r) & = &    { 1 \over  7-p } \, \ln \left[f_-(r) f_+(r) \right]
+{ (p-3)\over 64 } \left((p+1)c_1 -\frac{3-p}{4}c_3 \right)
  \, h(r) 
\nonumber \\
& &  + {p+1\over 16 } \, 
\ln\left[ \cosh( k \, h(r) ) 
- c_2\, \sinh( k \, h(r) ) \right],
\non
\phi(r) & = &    {(7-p)\over 16 }
\left((p+1)c_1 -\frac{3-p}{4}c_3 \right)\, h(r)   
\non
& & + { 3-p\over 4} \, \ln\left[ \cosh(k \, h(r) ) 
- c_2\, \sinh( k \, h(r) ) \right] , 
\non
e^{\Lambda(r) } & = & - \eta  (c_2^2-1)^{1/2} \,  {\sinh (k \, h(r) ) 
\over  \cosh( k \,  h(r) )  - c_2\, \sinh( k \, h(r) ) } \ ,
\label{solutionsappen}
\end{eqnarray}
where 
\begin{eqnarray}
f_\pm(r) &\equiv & 1\pm \left(\frac{r_0}{r}\right)^{7-p} \ ,
\non
h(r)  &=& \ln \left[\frac{f_-(r)}{f_+(r)} \right] \ , 
\non
k^2 & = & {2(8-p) \over 7-p }
- c_1^2 + \frac{1}{4}\left(\frac{3-p}{2}\ c_1 +
\frac{7-p}{8}\ c_3 \right)^2 - \frac{7}{16}\ c_3^2
 \ , 
\non
\eta & = & \pm 1 \ .
\label{definitions1}
\end{eqnarray}
The parameter $\eta$ describes whether we are measuring the
``brane'' charge or the ``antibrane'' charge of the system.

\end{document}